\documentclass[a4paper,fleqn,usenatbib]{mnras}
\usepackage{newtxtext,newtxmath}

\usepackage[T1]{fontenc}
\usepackage{ae,aecompl}

\usepackage{graphicx}	
\usepackage{amsmath}	
\usepackage{amssymb}	

\usepackage{bm}
\newcounter{num}
\numberwithin{equation}{section}

\title[Forecasts for WDM from weakly lensed SNe Ia]{Forecasts for warm dark matter from weakly lensed Type \Roman{num}a supernovae}

\author[M. Takeda et al.]{
Mei Takeda,$^{1}$\thanks{E-mail: m-takeda@astro.sc.niigata-u.ac.jp}
Ryuichiro Hada,$^{2}$\thanks{E-mail: ryuichiro.hada@ipmu.jp}
Ken-ichi Oohara,$^{1}$\thanks{E-mail: oohara@astro.sc.niigata-u.ac.jp}
Toshifumi Futamase,$^{3}$\thanks{E-mail: tof@cc.kyoto-su.ac.jp}
\\
$^{1}$Graduate School of Science and Technology, Niigata University, Niigata 950-2181, Japan\\
$^{2}$Kavli Institute for the Physics and Mathematics of the Universe, The University of Tokyo, Kashiwa, Chiba 277-8583, Japan\\
$^{3}$Department of Astrophysics and Meteorology, Kyoto Sangyo University, Kita-ku, Kyoto 603-8555, Japan
}

\date{Accepted XXX. Received YYY; in original form ZZZ}

\pubyear{2019}

\begin{document}
\label{firstpage}
\pagerange{\pageref{firstpage}--\pageref{lastpage}}
\maketitle

\begin{abstract}
We investigate the possibility to have a constraint on the mass of thermal warm dark matter (WDM) particle from the expected data of the {\it Wide Field Infrared Survey Telescope} ({\sl WFIRST}) survey if all the dark matter is warm.
For this purpose we consider the lensing effect of large scale structure based on the warm dark matter scenario on the apparent magnitude of SNe \Roman{num}a.
We use {\sc halofit} as non-linear matter power spectrum and the variance of PDF.
We preform a Fisher matrix analysis and obtain the lower bound of $m_{\rm WDM}>0.167$keV.
\end{abstract}




\section{Introduction}

Based on the various observations such as the cosmic microwave \citep{Planck2015,Planck2018} and the large scale structure of the universe \citep{SDSSDR15}, $\Lambda$ cold dark matter ($\Lambda$CDM) model is now regarded as the standard model of cosmology, namely, the universe is totally flat and consists of baryonic matter with $\Omega_{b,0}=0.049$, cold dark matter with $\Omega_{\rm CDM,0}=0.268$ and the dark energy $\Omega_{\rm DE,0}=0.683$ where $\Omega_{X,0}$ is the density parameter of the component $X$ at the present. 

However, recent more detailed observations and simulations pointed out some discrepancies between theoretical predictions and observational results at small scales, such as the missing satellite problem (e.g. \cite{Klypin1999}), the core-cusp problem (e.g. \cite{Moore1999c}) and the {\it too-big-to-fail} problem (e.g. \cite{Boylan211}). 
The missing satellite problem is that the number of DM subhalos which be predicted N-body simulation is much greater than the number of actually observed satellite galaxies of our galaxy and M31.
The core-cusp problem is that the $\Lambda$CDM model predicts DM halo has a cuspy density profile which diverges in the central region, but observation of the rotation curves of DM dominated galaxies show that central core density profile has a constant mass density.
The {\it too-big-to-fail} problem is that $\Lambda$CDM simulations show much denser central density than the actually observed the central density of massive dwarf galaxies around our galaxy. 

These problems seem to be serious difficulties for the cold dark matter scenario of structure formation and thus there are some proposals to solve these difficulties. Among them we here focus on the idea that the dark matter is not cold but warm (WDM) \citep{Colombi1996, Bode2001}.

The nature of WDM that can solve the above problems is its longer free streaming scale.
WDM behaves like CDM at large scales, but at small scales it has the effect like neutrino.
This effect depends on the WDM particle mass, therefore, it is important to find some method using available observational data to determine the mass.
As candidate of WDM particle, two possibilities are considered: thermal WDM and sterile neutrino.
In this paper we are concerned with thermal dark matter particle, described in \ref{subsec:WDMmodel}.
Many simulations have suggested that the required mass to solve above problems is of the order of 1keV.
Furthermore, there are some studies for the estimation of the mass of the WDM particle from various observations. 
For example, \cite{Irsic2017} shows $m_{\rm WDM}\geq$ 5.3keV (at 2$\sigma$ CL) from the Lyman $\alpha$ (Ly$\alpha$) forest data.
\cite{Inoue2015} shows $m_{\rm WDM}\geq$ 1.3keV (at 2$\sigma$ CL) from weak lensing effect by line-of-sight structures in QSO-galaxy quadruple lens systems using high-resolution $N$-body simulations. \citet{Markovic2011,Smith2011,Martins2018} study the possibility of constraining the mass from galaxy observations (such as cosmic shear power spectrum and galaxy angular power spectrum).

In this paper, we give another possible method to constrain the mass of the WDM particle by using the effect of weak lensing on the magnitude-redshift relation of SNe Type \Roman{num}a.
SNe Type \Roman{num}a is known as the cosmological "standard candles" and has played an important role to constrain some of the cosmological parameters (recent observation result: \cite{DES2019}).
Furthermore, there are many studies on the effect of weak lensing on the observation of SNe \Roman{num}a.(e.g. \cite{SmithM2014}).
Unlike other method, such as using galaxy clustering and the Ly$\alpha$ forest, the lensing effect on SNe \Roman{num}a does not suffer from bias problem.
Unlike galaxies, SNe are point sources and thus are directly affected by all matter along the line of sight.
These facts led \cite{SmithM2014} to conclude that the lensing effect is important for the next generation of surveys which will observe many SNe \Roman{num}a at high redshift.
The lensing effect on redshift has been studied in detail by \citet{Hada2016,Hada2018} and is used to constrain the cosmological parameters in the $\Lambda$CDM universe.
They used the probability distribution function (PDF) of the SNe \Roman{num}a magnitude which were derived from the convergence PDF.
In this work, using non-linear power spectrum applied to the pure WDM model, Fisher analysis is performed on $m_{\rm WDM}$ and $\Omega_{m}$ according to \citet{Hada2016,Hada2018}.

The outline of this paper is as follows.
In Section \ref{sec:pdfWDM}, we introduce the PDF for the apparent magnitude of SNe \Roman{num}a in the WDM model.
In Section \ref{sec:fisher}, we describe the analysis method and show our results.
Finally, we discuss and conclude in Section \ref{sec:last}.


\section{PDF in the WDM model}
\label{sec:pdfWDM}

In this section, we introduce the PDF of the apparent magnitude fluctuations of SNe \Roman{num}a from the non-linear matter power spectrum of the WDM model.
In Section \ref{subsec:WDMmodel}, we introduce the pure WDM model according to \citet{Colombi1996,Viel2005}.
Before explaining the transfer function of power spectrum for WDM model, we explain some detail of WDM modeling and WDM candidates, thermal WDM and sterile neutrino.
We show that there are three independent parameters in the model and is one-to-one correspondence between these candidates.
In Section \ref{subsec:pdfWDM}, we introduce the variance of the Gaussian PDF of the magnitude in the $\Lambda$CDM model and apply to the WDM model.


\subsection{Warm Dark Matter Model}
\label{subsec:WDMmodel}
The influence of WDM is the same as CDM at large-scale but different at small-scale.
The difference between WDM and CDM can be cleary seen in the matter power spectrum because their velocity dispersion at mater-radiation equality ($t_{\rm eq}$) are different and thus we can check their features through the matter power spectrum.
In this paper, we assumed WDM dominated universe, i.e. all DM ($\sim 26\%$) is WDM.
Therefore, we can express the power spectrum in the WDM models by applying a modified transfer function in the $\Lambda$CDM models. Before that, we will mention a little more about WDM candidates.

There are two candidates for WDM, thermal WDM and sterile neutrino.
The former can be considered as a generalization of case of massive neutrinos.
WDM particles decoupled earlier than the standard model neutrinos (\cite{Colombi1996,Viel2005}),
\begin{eqnarray}
\frac{T_{\rm WDM}}{T_{\gamma}} = \biggl( \frac{4}{11} \biggr)^{1/3} \biggl[ \frac{10.75}{g_{*}(T_D)} \biggr]^{1/3},
\end{eqnarray}
where the factor $(4/11)^{1/3}$ comes from the ratio between neutrino temperature after its decoupling and the photon temperature $T_{\gamma}$.
$T_D$ is the temperature of the universe, $g_{*}(T_D)$ is the effective number of degree of freedom of the WDM particles, and both are when WDM particle decoupled from other species.
On the other hand, according to \cite{Dodelson1994}, the sterile neutrino is not considered to be in thermal equilibrium, and is the right-handed component in the forth generation of neutrinos.
While sterile neutrino is produced from neutrino oscillations, assuming the constancy of $g_{*}$, the density parameters of both types of WDM candidates are described as follows
\begin{eqnarray}
\omega_{\rm WDM} = \Omega_{\rm WDM}h^2 =  \beta \left( \frac{m_{\rm WDM}}{0.094{\rm keV}} \right).
\end{eqnarray}
where $\beta = ( T_{\rm WDM}/ T_{\nu} )^3$ for the thermal WDM, where $T_{\nu}$ is neutrino temperature, $\beta =\chi$ for sterile neutrino with $\chi$ is an arbitrary normalization factor.
Thus we can take $\omega_{\rm WDM}, m_{\rm WDM}, and T_{\rm WDM}$ as independent parameters in the WDM model.
Further, when we use $\omega_{\rm WDM}$ and $m_{\rm WDM}/T_{\rm WDM}$ as parameters, it can be shown that the thermal WDM mass $m_{\rm thermal}$ and the sterile neutrino mass $m_{\rm sterile}$ are in one-to-one correspondence,
\begin{eqnarray}
m_{\rm sterile} = 4.43 \left( \frac{m_{\rm thermal}}{\rm keV} \right)^{4/3}
\left( \frac{0.25\times(0.7)^2}{\omega_{\rm WDM}} \right)^{1/3} {\rm keV}.
\end{eqnarray}
In this paper, WDM is treated as thermal WDM.

\begin{figure}
    \centering
    \includegraphics[width=\columnwidth]{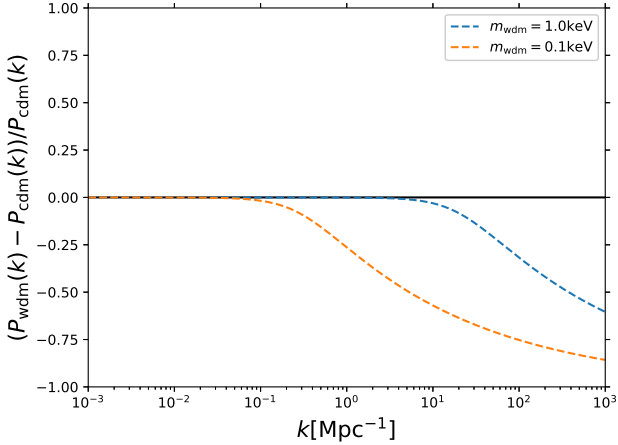}
    \caption{The ratio of non-linear matter power spectrum in the WDM models and in the $\Lambda$CDM models at $z=1$. The blue dashed line represents $m_{\rm WDM}=1$ keV and the orange dashed line represents $m_{\rm WDM}=0.1$keV. The lighter WDM mass more close to the $\Lambda$CDM model.}
    \label{fig:ratio}
\end{figure}

Next, we consider the matter power spectrum $P(k)$ in the WDM model.
The free streaming scale of WDM particle can be written as
\begin{eqnarray}
k_{\rm fs} = \frac{2 \pi}{\lambda_{\rm fs}}
\simeq 5  \left( \frac{m_{\rm WDM}}{\rm keV} \right) \left( \frac{T_{\nu}}{T_{\rm WDM}} \right){\rm Mpc}^{-1}.
\end{eqnarray}
The structures below this scale are suppressed.
This effect can be represented by the power spectrum of $\Lambda$CDM by applying the following modified transfer function.
\begin{eqnarray}
T(k) = \left[\frac{P_{\rm WDM}(k)}{P_{\Lambda {\rm CDM}}(k)} \right]^{1/2}.
\end{eqnarray}
For linear matter power spectrum, \cite{Bode2001} showed the fitting formula from full Boltzmann code calculation and \cite{Viel2005} revisited the best fit parameters.
They considered pure warm dark matter model in which our universe contains only warm dark matter.
Based on these, \cite{Viel2012} also determined the fitting formula for non-linear power spectrum as follows.
\begin{eqnarray}
\label{trans}
P^{\rm nonlin}_{\rm WDM} (k) &=& P^{\rm nonlin}_{\Lambda {\rm CDM}} (k) \left[  1 + (\beta k)^{\nu l} \right]^{-s/\nu},\\
\beta(m_{\rm WDM}, z) &=& 0.0476 \left( \frac{m_{\rm WDM}}{\rm keV} \right)^{-1.85} \left( \frac{1+z}{2} \right)^{1.3},
\end{eqnarray}
where $\nu=3, l=0.6$ and $s=0.4$.
As one can see in \cite{Viel2012}, when dealing with the non-linear power spectrum of the WDM model, one may think of halo model instead of the above fitting formula.
According to the comparison between the two approaches and simulation, the above fitting formula is more consistent with simulation than halo model at $k < 10h{\rm Mpc}^{-1}$.
Since this wave number range is important for considering the weak lensing effect, we used fitting formula using {\sc halofit} (\cite{Mead2016}) as the $P^{\rm nonlin}_{\rm CDM}$ in this paper.
In Fig.\ref{fig:ratio} we show the ratio of the non-linear power spectrum of the WDM models and the $\Lambda$CDM models.
Note, we use CAMB (\cite{Lewis2000}) for this calculation.

\subsection{PDF for SNe \Roman{num}a in the WDM model}
\label{subsec:pdfWDM}

The distance-redshift relation in an realistic inhomogeneous universe has been studied by many authors \citep[e.g.,][]{Dyper1972,Futamase1989, Okamura2009}.
The fluctuation of apparent magnitude of SN \Roman{num}a due to inhomogeneities in matter distribution of standard is linearly related to the fluctuation of matter density when the former fluctuation is sufficiently small.
This fact can be used to withdraw the information of matter distribution in the universe from SNe \Roman{num}a observations. On the other hand the matter distribution critically depends on the nature of DM. Thus it is reasonable to expect that a useful constraint for the nature of DM is obtained from the lensing observation of SNe \Roman{num}a.
In Section \ref{subsec:apamag}, \ref{subsec:CDMpdf}, We introduce our method according to \citet{Hada2016,Hada2018} and actually apply to the WDM model in Section \ref{subsec:apply}.


\subsubsection{The fluctuation of the apparent magnitude of SNe \Roman{num}a}
\label{subsec:apamag}

The apparent magnitude $m$ is defined by the flux $f$ from the light source as follows,
\begin{eqnarray}
    m = -2.5 \log_{10} f + {\rm const}.
\end{eqnarray}
We set the apparent magnitude actually observed as $m_{\rm obs}$, the one observed in the uniform isotropic universe as $m_{\rm true}$, and their difference as $\delta m$.
The factor of $\delta m$ is divided into two parts,
\begin{eqnarray}
    \delta m_{\rm tot} &=& \delta m_{\rm lens} + \delta m_{\rm othe}\nonumber \\
    &=& -2.5 \log_{10}\frac{f_{\rm lens}}{f_{\rm no-lens}} + \delta m_{\rm othe},
\end{eqnarray}
where $\delta m_{\rm tot} = \delta m$, $\delta m_{\rm lens}$ is the fluctuation due to lensing effect, and $\delta m_{\rm othe}$ is the fluctuation due to the others including an intrinsic ambiguity of the absolute magnitude.
Gravitational lensing is caused by the matter distribution between the light source and observer, therefore, it depends on the source redshift.
In general, the lensing effect is larger at higher-redshift.
On the other hand, we assume $\delta m_{\rm othe}$ does not have redshift dependence.
we express that $\sigma$ be the variance of the fluctuation, and $\sigma_{\rm othe}^2 = \sigma_{\rm means}^2 +\sigma_{\rm int}^2$ according to \cite{Hounsell2018}.
The first term on the right side expresses the dispersion of the distance precision per SN with $\sigma_{\rm means}$ which includes the accuracy of redshift measurement and light-curve fitting.
The second term comes from the intrinsic scatter of SNe \Roman{num}a with $\sigma_{\rm int}$ which reflects the fact that not all SNe follow distance-redshift relation.
Here, we use $\sigma_{\rm means}\simeq 0.08 {\rm mag}$, $\sigma_{\rm int} \simeq 0.09{\rm mag}$ according to \cite{Hounsell2018}, therefore, we got $\sigma_{\rm othe} \simeq 0.12 {\rm mag}$.


\subsubsection{PDF of apparent magnitude in $\Lambda$CDM models}
\label{subsec:CDMpdf}

In this paper, we assumed PDF of $\delta m_{\rm lens}$ and of $\delta m_{\rm othe}$ are both Gaussian distribution.
We follow \cite{Hada2018} to calculate the variance of the fluctuation of lensing effect.
The flux magnification $\mu=f_{\rm lens}/f_{\rm no-lens}$ due to the lens effect can be expressed using convergence $\kappa$ and shear $\gamma$,
\begin{eqnarray}
    \mu = \frac{1}{(1-\kappa)^2 - \gamma^2}.
\end{eqnarray}
From the result of high-resolution ray-tracing simulations by \cite{Takahashi2011}, we neglect the shear effect,
\begin{eqnarray}
    \mu \simeq (1 - \kappa)^{-2}.
\end{eqnarray}
Note that as convergence becomes larger, the approximation ignoring shear is worse.
Substituting this into $\delta m_{\rm lens}$,
\begin{eqnarray}
    \delta m_{\rm lens} &=& -2.5\log_{10}\mu \nonumber \\
    &\simeq& 5 \log_{10} |1-\kappa|.
\end{eqnarray}
Furthermore, we assumed the weak-lensing approximation such as $\kappa \ll 1$,
\begin{eqnarray}
    \delta m_{\rm lens} \simeq -\frac{5}{\ln 10}\kappa.
\end{eqnarray}
From the above relationship, the variance of the fluctuation of lensing effect becomes as follows,
\begin{eqnarray}
    \sigma_{\rm lens}^2 \simeq \left( \frac{5}{\ln 10} \right)^2 \langle \kappa^2\rangle.
\end{eqnarray}
The variance of the convergence is given by \cite{Bartelmann2001}
\begin{eqnarray}
    \label{kappav}
    \langle \kappa^2(z_s)\rangle&=&\frac{9}{8\pi}H_0^4 \Omega_{m0}^2 \int_0^{z_s} \frac{dz}{H(z)}(1+z)^2\left[ \frac{r(z)r(z,z_s)}{r(z_s)} \right] ^2 \nonumber \\
    &\ & \times \int_0^{\infty} d\ln k\ k^2 P_{{\rm nonlin}}(k,z),
\end{eqnarray}
where $z_s$ is redshift of light source, $r(z)$ is comoving distance, and $P_{{\rm nonlin}}$ is non-linear matter power spectrum.

The total variance is obtained by adding $\sigma_{\rm lens}^2$ and $\sigma_{\rm othe}^2$ because we consider Gaussian PDF, the expression of $\sigma_{\rm tot}$ is given by
\begin{eqnarray}
    \sigma_{\rm tot}^2(z_s) &=& \sigma_{\rm lens}^2(z_s) + \sigma_{\rm othe}^2
\end{eqnarray}


\subsubsection{Application to WDM models}
\label{subsec:apply}

\begin{figure}
    \centering
    \includegraphics[width=\columnwidth]{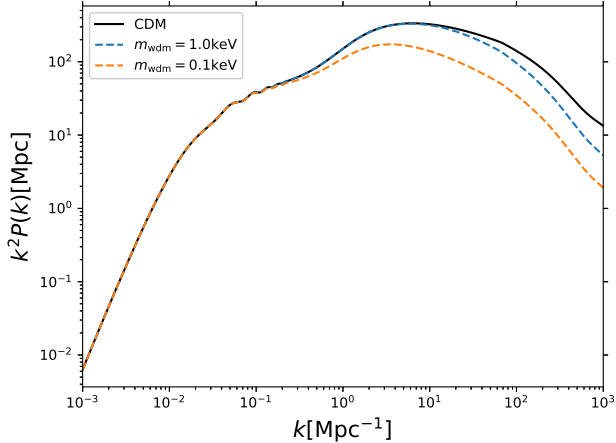}
    \caption{Comparison between the integrand $k^2 P(k)$ in the WDM model (color dashed line) and in the $\Lambda$CDM model (black solid line) at $z=1$. The blue dashed line show $m_{\rm WDM}=1$ keV and the orange dashed line show $m_{\rm WDM}=0.1$ keV. The heavier WDM mass more close to the $\Lambda$CDM model.}
    \label{k2p}
\end{figure}

\begin{figure}
    \centering
    \includegraphics[width=\columnwidth]{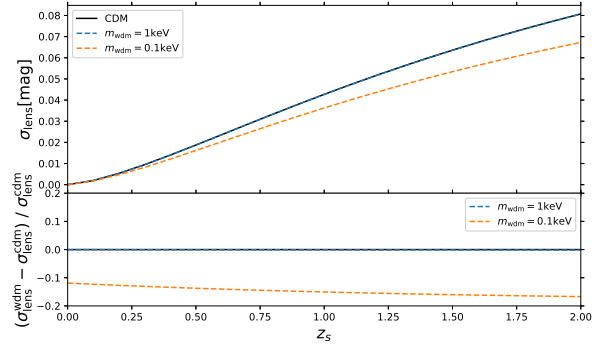}
    \caption{Top: Comparison between the $\sigma_{\rm lens}$ in the WDM model (color dashed line) and in the CDM model (black solid line). Down: The ratio of $\sigma_{\rm lens}^{\rm WDM}$ in the WDM model and in the $\Lambda$CDM model. The common horizontal axis $z_s$ is the redshift of the source. The blue dashed line show $m_{\rm WDM}=1$ keV and the orange dashed line show $m_{\rm WDM}=0.1$ keV. The behavior of $m_{\rm WDM}=1$ keV is almost the same as the $\Lambda$CDM model.}
    \label{sig}
\end{figure}

The application of the above expression to the WDM model is obtained by multiplying the transfer function of eq.(\ref{trans}) to the power spectrum of eq.(\ref{kappav}).
In Fig.\ref{k2p} and Fig.\ref{sig}, we showed $k^2 P(k)$ for $z=1$ and $\sigma_{\rm lens}$ for the $\Lambda$CDM model and the WDM model, respectively.

Next, we show how to select samples of SNe \Roman{num}a in this paper.
\citet{Hada2016, Hada2018} argued that the necessity of sample section and suggested that SNe, which passed through near galaxies core, may be eliminated, which corresponds to the upper bound of the wave number $k\simeq 10h^{-1}{\rm Mpc}$ in the non-linear power spectrum.
The chosen range of the wave number is the region the fitting formula and simulation coincide (\cite{Viel2012}) and at the same time one can eliminate the strong lensing events which is necessary because the above expressions of the lensing dispersion is derived under the approximation of the weak lensing.
In this paper we follow their sample section and set $k\simeq 10h^{-1}{\rm Mpc}$ as the maximum wave number.

\section{Forecasts for WDM particle mass}
\label{sec:fisher}

In this section, we explain our method and estimate the forecast of WDM particle mass with the expected data by {\sl WFIRST}.


\subsection{Setting cosmological parameter and SNe \Roman{num}a  data sets}
\label{subsec:parameter}

We used CAMB code for computing the matter power spectrum and cosmological parameters based on the results of Planck 2015 (\cite{Planck2015}) with $\Lambda$CDM model.
Their specific values are on Table \ref{para}.
The index $0$ means the parameter at the present $z=0$.
We assume that dark matter is warm dark matter: $\Omega_{\rm CDM,0} = \Omega_{\rm WDM}$.
From eq.(\ref{trans}), the case with $m_{\rm WDM}=\infty$ is identical with pure $\Lambda$CDM model and we then use it as a fidcial value.

\begin{table}
 \caption{Our fiducial parameters. They are given by Planck 2015 results with $\Lambda$CDM model. We changed CDM to WDM.}
 \label{para}
 \begin{center}
 \begin{tabular}{ccccccc}
  \hline
  $\Omega_{b,0}$ & $\Omega_{\rm CDM,0}$ & $\Omega_{k,0}$ & $\Omega_{m,0} $ & $\Omega_{\Lambda,0}$ & $\sum m_{\nu} [{\rm eV}]$ & $w$\\
  \hline
  0.049 & 0.267 & 0.0 & 0.317 & 0.682 & 0.06 & -1\\
  \hline
 \end{tabular}
 \end{center}
\end{table}

When we make the forecast, we prepare data sets (Table \ref{tab:wfirst}) of SNe \Roman{num}a which would be observed by {\it Wide Field Infrared Survey Telescope} ({\sl WFIRST})(\cite{Hounsell2018}).
{\sl WFIRST} is scheduled to launch in the mid-2020s, expected detect of the order of thousand of high redshift SNe \Roman{num}a up to $z=1.7$.
The difference between the WDM model and the $\Lambda$CDM model in the matter power spectrum is clearly seen in high redshift, therefore, we choose that survey plan as the next generation of space survey rather than ground survey such as Large Synoptic Survey Telescope (LSST)(\cite{LSST}).

\begin{table}
  \begin{center}
     \caption{Number of SN\Roman{num}a on each redshift bin $\Delta z = 0.1$ in {\sl WFIRST}}
      \label{tab:wfirst}
      \begin{tabular}{c|c} \hline
      redshift & number of SNe\Roman{num}a \\ \hline \hline
      $z = 0.2$ & $0.6 \times 10^2$   \\
      0.3 & $2.0 \times 10^2$   \\
      0.4 & $4.0 \times 10^2$   \\
      0.5 & $2.2 \times 10^2$   \\
      0.6 & $3.2 \times 10^2$   \\
      0.7-1.7 & $1.4 \times 10^2$ \\
       & (for each bin) \\ \hline
    \end{tabular}
  \end{center}
\end{table}


\subsection{Fisher's Information matrix}
\label{subsec:fisher}

Fisher information is useful tool in statistics when one wan to know how much informations on unknown parameters are obtained from the available data (e.g. \cite{Ly2017}).
We consider that the data $\bm{d}$ is random variable and follows the PDF $f_{\theta}(\bm{d})$ which has a parameter $\theta$.
The Fisher information is given from
\begin{eqnarray}
    I(\theta) \equiv -\left\langle \frac{\partial^2 \ln f_{\theta}(\bm{d})}{\partial \theta^2}\right\rangle
    = \left\langle \biggl( \frac{\partial \ln f_{\theta}(\bm{d})}{\partial \theta} \biggr)^2 \right\rangle,
\end{eqnarray}
 where $\langle \dotsi \rangle$ means ensemble average. 
 If there are some parameters $\bm{\theta}=\{ \theta_1,\dotsc,\theta_p \}$ ($p$ is the number of parameters), we can use $p\times p$ matrix having $(x, y)$ element
\begin{eqnarray}
    I_{xy}(\bm{\theta}) \equiv -\left\langle \frac{\partial^2 \ln f_{\bm{\theta}}(\bm{d})}{\partial \theta_x \partial \theta_y} \right\rangle
    = \left\langle \frac{\partial \ln f_{\bm{\theta}}(\bm{d})}{\partial \theta_x} \frac{\partial \ln f_{\bm{\theta}}(\bm{d})}{ \partial \theta_y} \right\rangle.
\end{eqnarray}
It is called Fisher's information matrix.
Here, according to \cite{Hada2018}, we express $f_{\bm{\theta}}(\bm{d})$ as follows
\begin{eqnarray}
    f_{\bm{\theta}}(\bm{d}) &=& \prod^N_{i=1} f_{\bm{\theta} ,i}\\
    f_{\bm{\theta} ,i} &=&  f_{\bm{\theta}}(m_{{\rm obs},i}).
\end{eqnarray}
We assume observation data sets are individual.
\begin{eqnarray}
    m_{{\rm obs}, i} = m_{\rm true}(z_i) + \delta m_{{\rm lens}, i} + \delta m_{{\rm othe}, i},
\end{eqnarray}
where $z_i, \delta m_{{\rm lens}, i}$ and $\delta m_{{\rm other}, i}$ are each redshift of $i$th apparent magnitude, magnitude fluctuation of lens and of other, respectively.
$m_{\rm true}(z_i)$ is the mean value, therefore, it is we decided by the standard luminosity distance in the Friedmann Robertson Walker metric.

\begin{figure*}
    \centering
    \includegraphics[width=18cm]{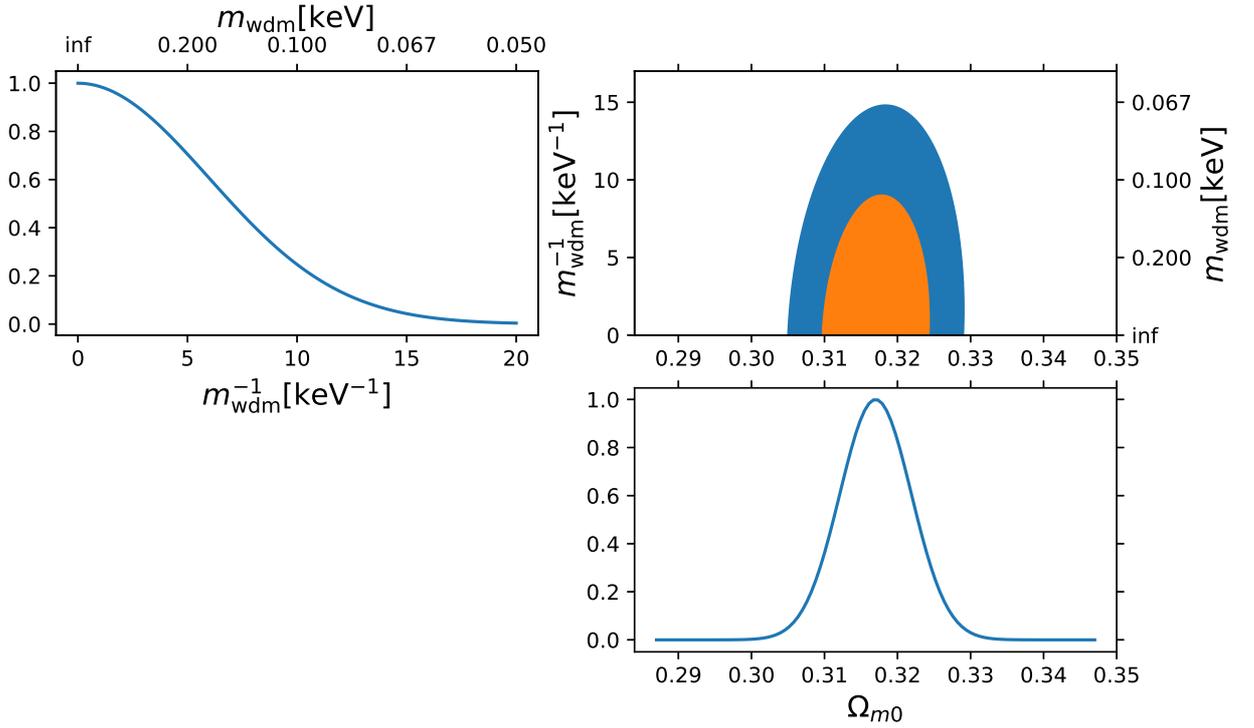}
    \caption{The result of the Fisher analysis of $\{ \Omega_{m0}, m_{\rm WDM}^{-1} \}$ from SNe \Roman{num}a for {\sl WFIRST}. The blue region show 2$\sigma$ (95\%) and the orange region show 1$\sigma$ (68\%). The parameters are the matter density at z=0 and Inverse of mass of warm dark matter. The memory of $m_{\rm WDM}$ is only carrying the value of the reciprocal of $m_{\rm WDM}^{-1}$ which variables actually used for calculation.}
    \label{fig:fisher}
\end{figure*}

We choose $\{ \Omega_{m0}, m_{\rm WDM}^{-1} \}$ as our parameters $\bm{\theta}$.
The inverse mass $m_{\rm WDM}^{-1}$ is more convenient because the fiducial value is $m_{\rm WDM}=\infty$ is difficult to handle.
In the actual calculation the fiducial value is chosen as $10^{-5}{\rm keV}^{-1}$ and follow the method developed by \cite{Markovic2011}.
The result is shown in Fig.\ref{fig:fisher} and thus $m_{\rm WDM} > 0.167$keV at $1\sigma$ is obtained as a minimal bound for the mass of WDM particle.
Although this value is smaller than the values obtained by other methods, it is important to note that our method is independent from other methods and free from bias.
We also discuss how to improve the constraint by our method. 
It is also important to note that if we do not observe the lensing effect in the m-z relation of SNe \Roman{num}a expected in $\Lambda$CDM model, it immediately reject Cold nature of DM.   

\section{Discussion and Conclusions}
\label{sec:last}

We investigated the forecast of constrain on the mass of the WDM particle from weakly lensed SNe \Roman{num}a for {\sl WFIRST}.
We assumed that the universe is dominated by thermal WDM with the non-linear matter power spectrum from obtained by multiplying a modified transfer function to the power spectrum in the $\Lambda$CDM model.
Also, we select SNe \Roman{num} samples by setting the wave number upper limit.

It turned out that these condition lead to $m_{\rm WDM} > 0.167$keV at $1\sigma$.
Our result include $m_{\rm WDM}\simeq1$keV which is required to solve the difficulties of CDM structure formation scenario, but is not severe compared with constrains obtained from galaxy clustering and the Ly$\alpha$ forest.
This is because the difference between $\sigma_{\rm lens}$ in the WDM model and in the $\Lambda$CDM model at $k < 10h {\rm Mpc}^{-1}$ is very small.
Therefore more accurate power spectrum for WDM is required to improve our constraint.
Also the improvement of $\sigma_{\rm othe}$ is essentially important to improve the constraint in our method.
In this respect, a huge low-redshift SNe \Roman{num}a sample expected by LSST may be used to establish a new statistical property to improve the intrinsic ambiguity of the absolute magnitude for SNe \Roman{num}a.
For example, $m_{\rm WDM} > 0.406$keV at $1\sigma$ when $\sigma_{\rm othe}\simeq0.06$ and $m_{\rm WDM} > 0.833$keV at $1\sigma$ when $\sigma_{\rm othe}\simeq0.01$.

In the present work we have used the Gaussian PDF of the convergence, but the PDF of density contrast in $\Lambda$CDM model is well approximated by a modified lognormal distribution \cite{Takahashi2011}, and thus the PDF of the convergence will be non-Gaussian.
Thus we expect to improve our constraint using non-Gaussian properties of PDF which will be our future work.

\section*{Acknowledgements}
This work is partly supported by a Grant-in-Aid for Science Research from JSPS (No.17K05453, No.18H04357 for T. F). RH is supported by JSPS Research Fellowships for Young Scientists (No.19J00513).




\bibliographystyle{mnras}
\bibliography{ref} 







\bsp	
\label{lastpage}
\end{document}